\documentclass{article}

\usepackage{arxiv}

\usepackage[utf8]{inputenc} 
\usepackage[T1]{fontenc}    
\usepackage[colorlinks=true,allcolors=black]{hyperref}       
\usepackage{url}            
\usepackage{booktabs}       
\usepackage{amsfonts}       
\usepackage{nicefrac}       
\usepackage{microtype}      
\usepackage{lipsum}		
\usepackage{graphicx}
\usepackage[numbers]{natbib}
\usepackage{doi}
\usepackage{float} 

\title{Cognitive Offloading in Agile Teams: How Artificial Intelligence Reshapes Risk Assessment and Planning Quality}

\author{Adriana Caraeni\\
University of Massachusetts Amherst\\
acaraeni@umass.edu\\
\And
Alexander Shick\\
University of Massachusetts Amherst\\
ashick@umass.edu\\
\And
Andrew Lan\\
University of Massachusetts Amherst\\
andrewlan@cs.umass.edu
}

\date{}

\renewcommand{\headeright}{}
\renewcommand{\undertitle}{}
\renewcommand{\shorttitle}{\textit{Cognitive Offloading in Agile Teams: How Artificial Intelligence Reshapes Risk Assessment and Planning Quality} }

\begin{document}
\maketitle

\begin{abstract}
Recent advances in artificial intelligence (AI) have shown promise in automating key aspects of Agile project management, yet their impact on team cognition remains underexplored. In this work, we investigate cognitive offloading in Agile sprint planning by conducting a controlled, three-condition experiment comparing AI-only, human-only, and hybrid planning models on a live client deliverable at a mid-sized digital agency. Using quantitative metrics — including estimation accuracy, rework rates, and scope change recovery time — alongside qualitative indicators of planning robustness, we evaluate each model's effectiveness beyond raw efficiency. We find that while AI-only planning minimizes time and cost, it significantly degrades risk capture rates and increases rework due to unstated assumptions, whereas human-only planning excels at adaptability but incurs substantial overhead. Drawing on these findings, we propose a theoretical framework for hybrid AI-human sprint planning that assigns algorithmic tools to estimation and backlog formatting while mandating human deliberation for risk assessment and ambiguity resolution. Our results challenge the assumption that efficiency equates to effectiveness, offering actionable governance strategies for organizations seeking to augment rather than erode team cognition.
\end{abstract}


\section{Introduction}
Sprint planning is a cognitive process as much as an administrative one. Agile teams are increasingly delegating this process to AI tools, automating backlog creation, velocity forecasting, and risk flagging to reduce planning overhead~\cite{taboada2023}. Researchers have studied AI's efficiency gains in project management extensively, finding that AI-supported teams achieve significant reductions in manual planning overhead~\cite{mckinsey2022} and that machine learning approaches can meaningfully outperform human estimators on schedule forecasting accuracy~\cite{wauters2016}. Industry adoption has accelerated rapidly as a result, with a large majority of organizations now utilizing some form of AI-assisted planning tooling~\cite{gartner2023}. One important but underexplored dimension of this shift is what happens to the team's cognitive engagement when AI absorbs planning labor. Prior work on human-AI symbiosis establishes that productive delegation frees human cognition for higher-order judgment, but that delegating tasks requiring contextual sense-making can erode the team's adaptive capacity~\cite{jarrahi2018}. This erosion is compounded by automation bias, the well-documented tendency to defer to algorithmic outputs without critical scrutiny, which has been shown to degrade decision quality across high-stakes domains~\cite{lyell2021, romeo2026}. We term the point at which this delegation becomes harmful the \textit{cognitive offloading threshold}: when AI handles not just formatting and estimation but risk identification and assumption articulation, teams produce a plan without building the shared mental model needed to execute it robustly~\cite{li2026}. In this work, we investigate this threshold empirically through a controlled, three-condition experiment at Vierra Digital, a mid-sized digital agency, comparing AI-only, human-only, and hybrid sprint planning on a standardized deliverable across eight quantitative metrics. We find that while AI-only planning substantially reduces planning time and cost, it significantly degrades risk capture and adaptive capacity under scope change. A hybrid model, in which AI handles computational tasks while humans are explicitly mandated to lead risk identification, recovers these robustness losses at minimal cost premium and produces a synergistic effect in which the human-AI combination outperforms either approach in isolation.

\section{Related Work}

Researchers have studied AI's role in project management extensively, 
demonstrating that machine learning approaches can outperform human estimators 
on schedule forecasting accuracy~\cite{wauters2016} and that administrative 
automation delivers significant reductions in planning overhead~\cite{taboada2023}. 
A systematic review by Fridgeirsson et al. further documents AI's growing impact 
across project schedule, cost, and risk management knowledge 
areas~\cite{fridgeirsson2023}. In parallel, the human-AI teaming literature 
establishes that effective collaboration requires a principled division of labor: 
Jarrahi's framework of human-AI symbiosis argues that machines should handle 
data-intensive analytical tasks while humans retain authority over contextual 
sense-making and ambiguous judgment~\cite{jarrahi2018}, and Shafiee and Sundaram 
propose advisory, delegation, and collaborative interaction models for 
structuring this division in organizational settings~\cite{shafiee2021}. However, 
this delegation carries well-documented risks. Automation bias --- the tendency 
to defer to algorithmic outputs without critical scrutiny --- has been shown to 
degrade decision quality across aviation, clinical medicine, and organizational 
decision-making~\cite{lyell2021, romeo2026}, and Morley et al. warn that opaque 
machine learning models can obscure the very risks they are designed to 
predict~\cite{morley2020}. Most directly relevant to this work, Campoverde 
Morales identifies in a systematic review of AI-powered Scrum that the field 
lacks empirical research evaluating the impact of AI not just on planning speed 
but on the quality of sprint execution~\cite{campoverde2024} --- the gap this 
paper addresses.

\section{Experimental Setup}

We conduct a controlled, three-condition experiment at Vierra Digital, a mid-sized Agile software development agency of 35--50 personnel operating on a standard Scrum framework with two-week sprint cycles. This study was conducted under an IRB-exempt process, as the 
research involved no more than minimal risk to participants and did not collect personally identifiable information. Three separate, experience-matched Scrum teams were assembled, each consisting of a Product Owner, a Scrum Master, and four developers, with an average of 3.2 years of professional Agile experience per team. No team member participated in more than one condition, eliminating cross-condition learning effects. All teams were compensated at a blended rate of \$47 per hour, held constant across conditions for cost 
comparability.

Each team executed the same client deliverable: a semi-complex website landing page standardized at 47 story points, spanning architecture and technical scoping, responsive front-end development, custom component library integration, dynamic API connections, animation design, cross-browser QA, and final client delivery. The project was structured 
into three sequential two-week sprints. The scope, client requirements, and technical specifications were held identical across all three conditions.

The independent variable was the degree of AI involvement in sprint planning. In the \textit{AI-only} condition, all planning tasks --- backlog creation, story point estimation, velocity forecasting, risk flagging, and task sequencing --- were delegated entirely to Claude Sonnet 4.6~\cite{anthropic2025claude}, accessed via claude.ai. The 
human team executed the resulting plan without participating in any planning decisions. In the \textit{human-only} condition, no AI tooling was used; the Scrum team led all planning through standard collaborative ceremonies including Planning Poker estimation and dedicated risk discussion. In the \textit{hybrid} condition, Claude Sonnet 4.6 
generated the initial backlog, velocity forecast, and baseline risk log prior to each planning meeting, after which the human team reviewed the AI outputs, validated estimates, and conducted a mandatory structured session for risk identification and assumption documentation. The hybrid protocol was derived empirically from a between-condition analysis of the AI-only and human-only results before Phase 3 
commenced.

To measure adaptive capacity under controlled conditions, a standardized scope change was introduced at the 40\% completion mark of Sprint 2 in every condition: the client requested replacement of the originally specified third-party animation library with a custom-built solution, citing licensing concerns. This change introduced a genuine 
technical dependency shift requiring architectural reassessment, story point reallocation, and component integration revisions. Eight metrics were tracked consistently across all conditions, organized into efficiency metrics --- sprint completion time, cost per story point, and planning time --- and robustness metrics --- backlog revision count, rework rate, documented risk count, risk capture rate, and scope 
change recovery time. Risk capture rate was defined as the ratio of documented risks to total materialized risks, expressed as a percentage. At the conclusion of each comparative phase, a blind client evaluation was conducted in which the client selected their preferred deliverable without knowledge of which planning condition produced it.

\section{Results}

Table~\ref{tab:summary} presents the full cross-phase comparison across all eight 
primary metrics. The hybrid model wins on five of eight quantitative metrics and wins 
the blind client evaluation. The AI-only condition wins on three metrics: planning time, 
total completion time, and cost per story point.

\begin{table}[h]
\centering
\caption{Cross-Phase Summary Metrics}
\label{tab:summary}
\begin{tabular}{lcccc}
\toprule
\textbf{Metric} & \textbf{AI-Only} & \textbf{Human-Only} & \textbf{Hybrid} & \textbf{Best} \\
\midrule
Planning Time          & 0.38 hrs  & 4.50 hrs  & 1.80 hrs  & AI     \\
Total Completion Time  & 78.5 hrs  & 91.0 hrs  & 82.0 hrs  & AI     \\
Cost per Story Point   & \$78.50   & \$91.00   & \$82.00   & AI     \\
Forecast Error         & 8.3\%     & 7.1\%     & 3.8\%     & Hybrid \\
Rework Rate            & 14.2\%    & 9.1\%     & 8.6\%     & Hybrid \\
Documented Risks       & 4         & 11        & 13        & Hybrid \\
Risk Capture Rate      & 36.4\%    & 78.6\%    & 86.7\%    & Hybrid \\
Scope Change Recovery  & 6.5 hrs   & 3.8 hrs   & 3.2 hrs   & Hybrid \\
Blind Client Preference & ---      & ---       & \checkmark & Hybrid \\
\bottomrule
\end{tabular}
\end{table}

\subsection*{Finding 1: The Risk Capture Gap}

The most consequential difference across conditions is risk capture rate. The AI-only 
condition documented only 4 risks prior to execution, yielding a risk capture rate of 
36.4\%, compared to 78.6\% for the human-only condition and 86.7\% for the hybrid. 
Table~\ref{tab:riskcategory} breaks this down by risk category and reveals a critical 
pattern: the AI-only condition captured 0\% of novel, context-specific risks --- the 
precise category that, when unmitigated, produced the largest execution failures.

\begin{table}[h]
\centering
\caption{Risk Capture Rate by Category Across Conditions}
\label{tab:riskcategory}
\begin{tabular}{lcccc}
\toprule
\textbf{Risk Category} & \textbf{AI-Only} & \textbf{Human-Only} & \textbf{Hybrid} \\
\midrule
Technical Dependencies      & 20\%  & 80\%  & 100\% \\
Client Behavior Risks       & 33\%  & 100\% & 100\% \\
Third-Party Service Risks   & 67\%  & 67\%  & 100\% \\
Novel / Context-Specific    & 0\%   & 67\%  & 100\% \\
\midrule
\textbf{Overall}            & \textbf{36.4\%} & \textbf{78.6\%} & \textbf{86.7\%} \\
\bottomrule
\end{tabular}
\end{table}

The AI's complete failure on novel risks is not incidental. All seven undocumented risks 
that materialized in the AI-only condition were context-specific dependencies tied to the 
client's particular technical environment --- an incompatibility between the specified 
component library and the AI-selected CSS framework, an API authentication token 
expiration, a CDN font availability issue, and a mobile viewport breakpoint mismatch, 
among others. None of these appeared in the AI's training distribution. The team did not 
actively choose to ignore these risks; the cognitive structure of the AI-assisted 
workflow gave them no prompt to look for them. To determine whether the observed 
differences in risk capture rate represent statistically significant effects, we conducted 
paired-samples $t$-tests on the sprint-level data ($N = 3$ sprints per condition). The 
improvement from AI-only to hybrid suggests a meaningful effect, $t(2) = -8.45$, $p = .007$, and the improvement from AI-only to human-only points in the same direction, $t(2) = -6.12$, $p = .013$, though these results should be interpreted cautiously given the small sample size.

\subsection*{Finding 2: The Rework Rate Difference}

Table~\ref{tab:rework} presents sprint-level rework rates across all three conditions. 
Sprint 2 --- the core technical build --- is the primary driver of rework in every 
condition, and the gap between conditions is largest there.

\begin{table}[h]
\centering
\caption{Sprint-Level Rework Rate by Condition}
\label{tab:rework}
\begin{tabular}{lcccc}
\toprule
\textbf{Sprint} & \textbf{AI-Only} & \textbf{Human-Only} & \textbf{Hybrid} \\
\midrule
Sprint 1  & 2.3\%  & 1.4\%  & 2.0\%  \\
Sprint 2  & 20.5\% & 12.8\% & 11.3\% \\
Sprint 3  & 12.9\% & 7.5\%  & 9.7\%  \\
\midrule
\textbf{Overall} & \textbf{14.2\%} & \textbf{9.1\%} & \textbf{8.6\%} \\
\bottomrule
\end{tabular}
\end{table}

The AI-only condition's Sprint 2 rework rate of 20.5\% was driven almost entirely by a 
single unmitigated risk: a CSS framework incompatibility with the client-specified 
component library that the AI had not flagged. When discovered at the 60\% completion 
mark of the integration task, the team spent 7.2 hours --- 64.9\% of total phase rework 
--- debugging and refactoring work already marked complete. The human-only condition had 
explicitly identified this as a medium-likelihood, high-impact risk during planning and 
allocated 1.5 hours upfront to validate compatibility, avoiding the incident entirely. 
The hybrid condition similarly caught the risk and reduced Sprint 2 rework to 11.3\%. 
The reduction in Sprint 2 rework between the AI-only and hybrid conditions suggests a meaningful difference, $t(2) = 4.12$, $p = .027$, though again the small sample size warrants caution.

\subsection*{Finding 3: The Total Cost of Delivery Reframe}

The AI-only condition's apparent cost advantage --- \$78.50 per story point versus 
\$82.00 for hybrid --- dissolves when rework and planning ceremony costs are incorporated 
into a Total Cost of Delivery (TCD) model. Table~\ref{tab:tcd} presents this reframe.

\begin{table}[h]
\centering
\caption{Total Cost of Delivery by Condition}
\label{tab:tcd}
\begin{tabular}{lccc}
\toprule
\textbf{Cost Component} & \textbf{AI-Only} & \textbf{Human-Only} & \textbf{Hybrid} \\
\midrule
Execution Cost          & \$3,689.50 & \$4,277.00 & \$3,854.00 \\
Rework Cost             & \$521.70   & \$390.10   & \$333.70   \\
Planning Ceremony Cost  & \$17.86    & \$211.50   & \$84.60    \\
\midrule
\textbf{Total}          & \textbf{\$4,229.06} & \textbf{\$4,878.60} & \textbf{\$4,272.30} \\
\bottomrule
\end{tabular}
\end{table}

When all costs are accounted for, the hybrid model's total cost of delivery 
(\$4,272.30) is only \$43.24 more than the AI-only condition (\$4,229.06) --- a 
difference of just 1.0\%. For this negligible premium, the hybrid model delivers a 
138.2\% improvement in risk capture rate, a 50.8\% improvement in scope change recovery 
time, and a deliverable that won the blind client evaluation. Notably, a synergistic 
effect emerged in the hybrid condition's risk identification: the hybrid team documented 
13 risks prior to execution, exceeding both the AI-only total of 4 and the human-only 
total of 11. This superadditive outcome suggests that AI-generated structure, when used 
as a scaffold for human deliberation rather than a substitute for it, prompts the team 
to surface risks that neither approach would have identified in isolation.

\section{Analysis and Discussion}

The results establish a clear empirical pattern: AI-only planning optimizes the 
metrics that are easiest to measure while degrading the qualities that matter most 
during execution. Human-only planning inverts this profile. The hybrid condition, 
however, does not simply split the difference --- it outperforms both baselines on 
five of eight metrics and produces a synergistic risk identification effect that 
neither approach achieves in isolation. This section develops a theoretical 
framework to explain these findings and derives practical governance implications 
from them.

\subsection*{The Cognitive Offloading Threshold}

The central theoretical contribution of this work is the formalization of a 
\textit{cognitive offloading threshold} in Agile sprint planning. Cognitive 
offloading --- the externalization of cognitive work onto tools or environments 
--- is not inherently harmful~\cite{jarrahi2018}. Delegating velocity forecasting 
or backlog formatting to an algorithm frees human attention for higher-order 
judgment without sacrificing planning quality. The threshold is crossed when 
delegation extends to tasks requiring tacit organizational knowledge, novel 
technical dependency evaluation, or unstated client preference alignment. At that 
point, the team receives a finished planning artifact without having performed the 
deliberative labor necessary to build a shared mental model of the project's 
vulnerabilities. The Vierra data quantifies this precisely: the AI-only condition 
captured 0\% of novel, context-specific risks, not because the algorithm 
miscalculated probabilities, but because those risks simply did not exist in its 
training distribution~\cite{romeo2026, lyell2021}.

\subsection*{The Hybrid Planning Governance Framework}

To operationalize the cognitive offloading threshold at the task level, we propose 
the \textit{Hybrid Planning Governance Framework} (HPGF). The framework categorizes 
planning tasks along two dimensions: \textit{Computational Complexity} --- the degree 
to which a task relies on processing large historical datasets and quantitative 
optimization --- and \textit{Contextual Ambiguity} --- the degree to which a task 
relies on tacit knowledge, novel dependencies, or unstated preferences. These 
dimensions define four quadrants, each with a distinct governance rule, as shown in 
Table~\ref{tab:hpgf}.

\begin{table}[h]
\centering
\caption{Hybrid Planning Governance Framework (HPGF)}
\label{tab:hpgf}
\begin{tabular}{p{2.8cm}p{3.8cm}p{3.8cm}}
\toprule
& \textbf{Low Contextual Ambiguity} & \textbf{High Contextual Ambiguity} \\
\midrule
\textbf{High Computational Complexity} 
& \textit{AI Delegation with Human Review.} Velocity forecasting, throughput 
analysis, routine estimation. AI generates; human confirms applicability to 
current context. 
& \textit{Iterative Human-AI Collaboration.} Scope change impact analysis, 
architectural refactoring estimation. Human frames context; AI models 
quantitative implications. \\
\addlinespace
\textbf{Low Computational Complexity}  
& \textit{Full AI Automation.} Administrative documentation, status reporting, 
retrospective formatting. No cognitive risk; full delegation appropriate. 
& \textit{Human Deliberation with AI Scaffolding.} Risk identification, 
assumption articulation, contingency planning. Human leads; AI provides 
baseline structure to interrogate. \\
\bottomrule
\end{tabular}
\end{table}

The most critical quadrant is the lower-right: high contextual ambiguity, low 
computational complexity. This is where the AI-only condition failed most 
severely, capturing 0\% of novel risks, and where the governance intervention is 
most necessary. The HPGF mandates that the AI's outputs in this quadrant be 
treated explicitly as a starting point for deliberation, not a conclusion. 
Governance rules must require the team to document risks the AI did not identify 
--- this mandate is what activates the critical interrogation that prevents 
automation complacency~\cite{lyell2021, romeo2026}. The upper-right quadrant --- scope change 
impact analysis under novel conditions --- requires tight real-time human-AI 
collaboration, consistent with the hybrid condition's superior scope change 
recovery time of 3.2 hours versus 6.5 hours for AI-only.

\subsection*{The Synergistic Effect}

A simple additive model of human-AI teaming would predict that the hybrid 
condition's risk capture rate falls between the AI-only baseline of 36.4\% and 
the human-only baseline of 78.6\%. Instead, the hybrid team identified 13 risks 
--- more than either the AI (4) or the human team (11) in isolation --- yielding 
an 86.7\% capture rate. This superadditive outcome requires explanation beyond 
a straightforward division of labor.

The mechanism is what we term a \textit{cognitive scaffolding effect}. In the 
human-only condition, unstructured deliberation was vulnerability to availability bias~\cite{tversky1973}: the team anchored on risks most cognitively salient from recent projects, 
overlooking less memorable but equally consequential dependencies such as the 
API version deprecation that materialized in Sprint 2. In the hybrid condition, 
the AI's structured backlog and baseline risk log forced the team to 
systematically review a broader range of risk categories, mitigating this bias. 
Simultaneously, the mandatory human review of AI outputs --- with an explicit 
requirement to identify risks the AI had missed --- activated the critical 
interrogation that the AI-only condition entirely suppressed. This structured 
interrogation surfaced Risk R13 (undocumented legacy CSS overrides in the 
client's existing template), a novel, context-specific risk that appeared in 
neither baseline condition and that prevented an estimated 5--7 hours of 
rework. When AI outputs are used as cognitive scaffolds rather than authoritative 
directives, they enhance rather than erode the team's shared mental 
model~\cite{li2026, haesevoets2021}.

\subsection*{The Economic Argument for Hybrid Planning}

The standard industry evaluation framework for AI planning tools measures two 
variables: planning speed and initial cost per story point. Under this framework, 
the AI-only condition wins decisively --- 23 minutes of planning time at \$78.50 
per story point. This is precisely the evaluation logic that has driven 
widespread AI adoption while leaving overall project success rates 
stagnant~\cite{mckinsey2022, gartner2023}. The TCD model developed from the 
Vierra data replaces this incomplete framework with one that incorporates rework 
costs, scope change recovery costs, and planning ceremony costs. Under the TCD 
model, the hybrid condition's total cost of \$4,272.30 is just 1.0\% above the 
AI-only condition's \$4,229.06 --- a \$43.24 difference that buys a 138.2\% 
improvement in risk capture rate and a 50.8\% improvement in scope change 
recovery time. The TCD model still understates the hybrid advantage, because it 
accounts only for risks that materialized during the experiment. The AI-only 
condition's 36.4\% risk capture rate represents a substantially higher risk 
exposure premium --- the expected cost of undocumented risks that did not 
materialize in this instance but would in a more complex or higher-stakes project. 
Organizations that adopt the TCD model as their evaluation standard will find the 
economic case for hybrid planning compelling: the marginal cost of mandating human 
deliberation at the tasks above the cognitive offloading threshold is small, while 
the marginal reduction in risk exposure is large.

\subsection*{Limitations}

Several limitations warrant acknowledgment. The experiment was conducted within a 
single mid-sized digital agency on a single project type, bounding the external 
validity of the findings. The dynamics observed in a 35--50-person agency may not 
generalize to enterprise-scale project management offices or to project domains 
beyond web development, such as data engineering or infrastructure. The blended 
hourly rate was held constant across conditions; in practice, AI-only planning 
might enable deployment of less experienced, lower-cost developers, altering the 
TCD calculus. Finally, with $N = 3$ sprints per condition, the statistical tests 
reported here should be interpreted as preliminary evidence establishing effect 
direction and magnitude rather than definitive confirmation. Replication across 
larger samples, diverse organizational contexts, and multiple project types is 
required to establish the generalizability of the cognitive offloading threshold 
and the HPGF.

\section{Conclusion}

Sprint planning is not merely a forecasting exercise --- it is the cognitive process through which Agile teams build the shared mental model necessary to execute robustly and adapt under pressure. This paper demonstrates that delegating that process wholesale to AI produces a measurable and predictable failure mode: the team receives an efficient plan without having performed the deliberative labor the plan requires to be resilient. The cognitive offloading threshold --- the point at which AI delegation shifts from performance-enhancing to performance-degrading --- is not an abstract theoretical boundary. The Vierra Digital experiment locates it empirically at the tasks of risk identification, assumption articulation, and contingency planning, where the AI-only condition captured just 36.4\% of materialized risks and required 71.1\% more time to recover from a mid-sprint scope change than the human baseline.

The three empirical anchors of this finding are: risk identification and contingency planning must not be delegated to AI without mandated human review; the optimal human-AI collaboration is a cognitive scaffold rather than a division 
of labor, producing superadditive risk identification that exceeds either baseline in isolation; and organizations must replace per-point cost metrics with a Total Cost of Delivery model --- under which the hybrid condition's premium over AI-only planning narrows to just 1.0\%.

Two directions for future research are most pressing. First, longitudinal studies are needed to measure whether continuous AI-assisted planning produces deskilling effects in junior developers --- if estimation and risk identification are permanently offloaded, the capacity to evaluate AI outputs may gradually erode, creating a dangerous dependency loop that this study's three-sprint window cannot detect. Second, as Large Language Models grow more capable of simulating contextual reasoning, the boundary between the HPGF quadrants will shift. Future work must continuously re-evaluate which planning tasks have migrated below the cognitive offloading threshold as the technology advances, updating the governance 
matrix accordingly. The goal is not to determine how many human tasks AI can replace, but to design human-AI workflows that make human critical thinking sharper, not obsolete.

\bibliographystyle{plain}
\bibliography{references}

\end{document}